\documentclass{llncs}

\usepackage{algorithmic}
\usepackage{algorithm}
\usepackage{setspace}
\usepackage{url}
\usepackage{subfigure}
\usepackage{amsmath}
\usepackage{amssymb}
\usepackage{geometry}
\usepackage{graphics}
\usepackage{graphicx}
\usepackage{breqn}

\begin{document}

%\title{Realize Multi-party Logistic Regression Learning under Differential Privacy}
\title{Privacy-Preserving Multiparty Learning For Logistic Regression }

\author{$^\S$Wei Du\thanks{This work was done when Wei Du was at the University of Arkansas.},   $^\dag$Ang Li, $^\dag$Qinghua Li}
\institute{$^\S$Department of Electrical and Computer Engineering, Michigan State University\\$^\dag$Department of Computer Science and Computer Engineering, University of Arkansas\\
	Email:  $^\S$duwei1@msu.edu, $^\dag$\{angli,  qinghual\}@uark.edu}

\maketitle

\begin{abstract}
In recent years, machine learning techniques are widely used in numerous applications, such as weather forecast, financial data analysis, spam filtering, and medical prediction. In the meantime, massive data generated from multiple sources further improve the performance of machine learning tools. However, data sharing from multiple sources brings privacy issues for those sources since sensitive information may be leaked in this process. In this paper, we propose a framework enabling multiple parties to collaboratively and accurately train a learning model over distributed datasets while guaranteeing the privacy of data sources. Specifically, we consider logistic regression model for data training and propose two approaches for perturbing the objective function to preserve $ \epsilon $-differential privacy. The proposed solutions are tested on real datasets, including Bank Marketing and Credit Card Default prediction. Experimental results demonstrate that the proposed multiparty learning framework is highly efficient and accurate. 
\end{abstract}

\section{Introduction}
The past few decades have witnessed an increasing role that machine learning techniques play in both academic and industry communities. 
These techniques can be widely used to extract useful information from datasets in various fields \cite{kotsiantis2007supervised}. At the same time, the advent of the big data era provides a better platform for its further development. For example, some advertisement companies collect massive data from social media, such as search history from Google and individuals' interactions from Facebook, and analyze the data to lock in targeted customers and improve accuracy of posting advertisements. Machine learning algorithms also have applications in the medical area. Taking cancer scan for instance, some types of cancers are really difficult for even experienced doctors to accurately determine cancer staging, but it has been reported that intelligent computers can help do this with higher accuracy. The combination of data analytics and efforts by doctors can give better medical treatment plans. Machine learning also exerts its effect in the field of finance. Financial companies accumulate a large amount of data records of their customers including purchase history, credit level, and loans and mortgage repayment activities. Then these companies could develop automatic, intelligent fraud detection systems to actively learn the behaviors of customers and distinguish potential threats and bankrupt cases, which can significantly reduce asset loss and bad debts. 

A typical machine learning paradigm is to make targeted predictions based on a single dataset \cite{witten2016data}. However, the data resources are increasingly distributed and stored by different owners. For example, medical data can be distributed in several hospitals and healthcare institutes; personal credit history, asset status, and accounting information can be distributed across multiple financial companies. Combining data from multiple sources for learning can usually derive better prediction performance. As a result, the traditional paradigm of learning from a single dataset is experiencing transitions towards collaborative learning, i.e., data from multiple parties are used to collaboratively train a learning model. A conventional collaborative learning approach is to have a central party, i.e., a virtual server and let multiple data owners directly upload their data to the server for training \cite{bouwen2004multi,pathak2010multiparty,rajkumar2012differentially,ohrimenko2016oblivious}. 

Although collaborative learning achieves better performance than singe dataset based learning, concerns on privacy are arising. It is possible that during the process, the private information of each party, e.g. health data records, can be disclosed, which will cause privacy leakage. Furthermore, the privacy leakage issue will cause mistrust between participating parties, preventing them from sharing their data to the central server. Therefore, it becomes increasingly important to design a protocol to train a learning model from the datasets from multiple parties, while at the same time preserving their privacy.

Some work has been done for privacy-preserving machine learning in the past few years, and those protection techniques can be mainly divided into three groups. The first group perturbs the original data with randomized algorithms \cite{kabir2007data,liu2006random,shobana2012deriving}. Although the perturbation techniques can protect confidential information, perturbed data may differ from the original data to a large extent, and thus decrease the accuracy of the resulted training model. Secondly, anonymization is also a popular method to protect users’ sensitive information \cite{inan2009using}. For example, we can remove the name and identification number from individual history records. Then the accuracy of the training model will not be affected and the privacy of data owners can also be preserved. However, such anonymization techniques are vulnerable to attacks involving auxiliary information. Thirdly, some cryptographic techniques have also been reported in previous studies \cite{bos2014private,graepel2012ml}. In \cite{graepel2012ml}, Thore \textit{et al} reported a homomorphic encryption scheme to retain both the privacy of training and testing examples. However, homomorphic encryption techniques will incur intensive computations, making it impractical for large-scale applications. 

Recently, differential privacy \cite{dwork2008differential} as an arising notation has attracted much attention in the field of privacy. Theoretically, it could offer formal privacy guarantees no matter what auxiliary information the attackers have. There has been a lot of work on the realization of machine learning models under differential privacy. A typical approach is first generating noise via Laplace mechanism or exponential mechanism and then building a noisy model for their dataset using these generated noises \cite{dwork2008differential}. Some other approaches modify the objective function of the training model \cite{dwork2014algorithmic}. These mechanisms can perturb the objective function by adding noise, and output predictions of the perturbed noisy model. However, most of them are focusing on the single-party setting, and have not applied to multiparty learning.

In this paper, we propose a framework for privacy-preserving multiparty learning among multiple data owners. The proposed framework achieves differential privacy, providing theoretical guarantees for each party’s privacy. The framework focuses on the application of logistic regression for model building, but can be easily applied or extended to other machine learning models. In this framework, each data owner first trains its model locally, and each of them will obtain an output objective function for their local model. We design two approaches of noise generation in this process to meet differential privacy. Then these data owners will upload their learned local parameters to a central server for sharing. The central server will average the uploaded local parameters and send the averaged parameters back to those data owners. Local data owners will then incorporate the averaged parameters to retrain their local model. The above process will be repeated iteratively until the parameters converge. The proposed framework is based on the weighted parameter averaging mechanism since the number of data records for different data owners might be different. Experimental results show that the proposed framework is computationally efficient. The main contributions of this paper are summarized as follows:
\begin{itemize}
	\item We propose a differentially private framework for multiple parties to collaboratively build learning models using logistic regression, which provides theoretical privacy guarantees for those parties.  
	\item Two efficient mechanisms are designed for generating noise during local learning to achieve differential privacy for multiparty learning. 
	\item We propose a weighted parameter sharing mechanism for multiple data owners with different sizes of data records.
	\item We run extensive experiments to evaluate the performance of the proposed framework using real datasets, and the results show high efficiency and accuracy of the proposed solution. 
\end{itemize}

The rest of the paper is organized as follows. Section II introduces preliminary background knowledge. Section III elaborates the two different approaches for realizing differentially private learning in a multiparty setting. In addition, theoretical privacy analysis for the proposed mechanism is also provided. Section IV presents evaluation results. Section V discusses related work and their difference from our work, and Section VI concludes the paper.

\section{Preliminaries}
\subsection{Differential Privacy}
Differential privacy is an important concept in the area of privacy. It formally guarantees that no matter what change has been made to any particular element in a database an attacker cannot tell the difference in the output of a randomized algorithm [13]. 

\indent \textit{Definition 1.} A randomized algorithm $ \mathcal{M} $ that takes the elements in $ \mathcal{D} $ and outputs a function $ \mathcal{M(D)} $ achieves $ \epsilon$-differential privacy if
\begin{equation}
\dfrac{P(\mathcal{M(D)} \in S)}{P(\mathcal{M(D')} \in S)} \leq e^{\epsilon}
\end{equation}
where $ S $ is the output range of $ \mathcal{M(D)} $. $ \mathcal{D} $ and $ \mathcal{D'} $ are a pair of neighborhood databases differing in a single item. $ \epsilon $ is a privacy budget that controls the strength of the privacy of the algorithm $ \mathcal{M} $ for any pair of neighborhood databases. For smaller $\epsilon$, the output  $ \mathcal{M(D)} $ is almost the same for any  pair $ \mathcal{D} $ and $ \mathcal{D'} $, making it hard for the adversarial party to identify the difference between these two neighborhood databases. To make it more clear, we have: 
\begin{equation}
e^{-\epsilon}\leq \dfrac{P(\mathcal{M(D)} \in S)}{P(\mathcal{M(D')} \in S)} \leq e^{\epsilon}
\end{equation}
The derivation of Eq. (2) is based on the interchangeability of the \textit{Definition 1}. For smaller $\epsilon$, we have the approximation formula:
\begin{equation}
e^{\epsilon} = 1 - \epsilon
\end{equation}
The combination of Eq. (2) and Eq. (3) will give:
\begin{equation}
1 - \epsilon \leq \dfrac{P(\mathcal{M(D)} \in S)}{P(\mathcal{M(D')} \in S)} \leq 1 + \epsilon
\end{equation}
Since $ 0 < P(\mathcal{M(D)} \in S), P(\mathcal{M(D')} \in S) < 1$, the expansion of Eq. (4) can be rewritten as:
\begin{equation}
|P(\mathcal{M(D)} \in S) -  P(\mathcal{M(D')} \in S)| < \epsilon
\end{equation}
It can be clearly seen from Eq. (5), if $ \epsilon $ is a negligible variable, any malicious party inquiring $ \mathcal{M(D)} $ cannot distinguish $ \mathcal{D} $ and $ \mathcal{D'} $. Then strong privacy of the data owner will be achieved.

\indent Several methods have been used to satisfy $ \epsilon $-differential privacy and Laplace mechanism is a commonly used one \cite{dwork2008differential}. In particular, the Laplace mechanism adds a random noise to the output of an algorithm, where the random noise is drawn from Laplace distribution depending on the global sensitivity of the algorithm \cite{dwork2014algorithmic}. 

\indent \textit{Definition 2.} Given a real-valued vector mapping function: $ \mathcal{D} \to \mathcal{M(D)}$, the sensitivity of $ \mathcal{M(D)}$ is denoted as: 
\begin{equation}
\Delta S_{\mathcal{M}} = \text{max} \Arrowvert \mathcal{M(D)} - \mathcal{M(D')} \Arrowvert_1
\end{equation}
where $ \mathcal{D} $ and $ \mathcal{D'} $ are any pair of neighborhood databases, and $ \Arrowvert \mathcal{M(D)} - \mathcal{M(D')} \Arrowvert_1 $ is the $ l_1 $ distance between $ \mathcal{M(D)} $ and $ \mathcal{M(D')} $. The sensitivity $ \Delta S_{\mathcal{M}}  $ describes the maximum variation of the output $ \mathcal{M(D)}$ under single-item changes.

\indent \textit{Definition 3.} The Laplace distribution centered at $ \mu $, and scaled with $ b $ has the following probability density function: 
\begin{equation}
Lap(x|\mu, b) = \dfrac{1}{2b}exp(-\dfrac{|x - \mu|}{b})
\end{equation}
Then to compute the noisy output on database $ \mathcal{D} $, we will have:
\begin{equation}
\mathcal{M'(D)} = \mathcal{M(D)} + Lap(x|0, \dfrac{\Delta S_{\mathcal{M}}}{\epsilon})
\end{equation}
where $ Lap(x|0, \dfrac{\Delta S_{\mathcal{M}}}{\epsilon}) $ indicates that the Laplace distribution follows zero mean and $ \dfrac{\Delta S_{\mathcal{M}}}{\epsilon} $ scale. The noise generation mechanism in Eq. (8) can provide $\epsilon$-differential privacy for any randomized algorithm $ \mathcal{M} $.

\subsection{Linear Regression and Logistic Regression}
Consider a basic machine learning task for binary classification, where a database $ \mathcal{D} $ is given. Suppose $ \mathcal{D} $ consists of $ N $ samples $ X_1, X_2, X_3, \cdots, X_n $. Each sample $ X_i $ is denoted as  ($ x_{i1} $, $ x_{i2} $, $ x_{i3} $,$  \cdots $, $ x_{im} $, $ y_i$) , where $ x_{i1}, x_{i2}, x_{i3}, \cdots, x_{im}$  are the $ m $ attributes of sample $ X_i $, and $ y_i \in \{0, 1\}$ is the corresponding binary label. For example, when we record cancer history of patients, we need to write down some basic information, such as age, weight, height, and so on. In addition, we also need some high-level information, including blood pressure, heart rate, and medical related indexes. All these information corresponds to the $ m $ attributes of a data record. The label $ y_i \in \{0, 1\}$ indicates existence of a disease. A patient will be labeled as $ 1 $ if diagnosed with a particular cancer, and otherwise will be labeled as $ 0 $. 

\indent From the perspective of machine learning scientists, it is assumed that there exists a hidden relationship between the $ m $ attributes $ x_{i1}, x_{i2}, x_{i3}, \cdots, x_{im} $ and the recorded label $ y_i \in \{0, 1\}$. The objective of a machine learning algorithm is to learn this particular relationship which empowers us to predict the label of a data record as accurate as possible given the corresponding attributes logged. Suppose the prediction function $ f $ taking the input of $i$th element $ (\textbf{x}_i:x_{i1}, x_{i2}, x_{i3}, \cdots, x_{im} )$, and outputting the predicted label $ \hat y_i $ is expressed as the following:
\begin{equation}
\hat y_i = f(\textbf{x}_i)
\end{equation}
The above function $ f $ is usually formulated as an optimization problem that one would like to maximize the prediction accuracy. The number of errors made by the prediction function $ f $ over the entire dataset is shown as:
\begin{equation}
\sum_{n = 1}^{N}\Pi_{\hat y_i \neq y_i}
\end{equation}
where $ \Pi $ is a loss function that evaluates the difference between the predicted label and the real label. Our goal is to minimize the number of prediction errors, and the optimization problem is stated as \cite{kutner2004applied}:
\begin{equation}
\textbf{w}^* = \text{arg min}_{\textbf{w}}\sum_{n = 1}^{N}\Pi_{\hat f(\textbf{x}_i) \neq y_i}
\end{equation}
where $ \textbf{w} $ is the parameter of the function $ f $, and $ \textbf{w}^* $ is the optimal result of Eq. (11). 
\\ \indent The types of objective function and loss evaluation could vary. In this paper, we focus on the application of logistic regression, which is a commonly used classification technique. In the following part, we will first introduce linear regression and then extend to logistic regression.
 
\indent Linear regression is the basic regression model, and the prediction function $ f $ is assumed to be linearly dependent on the $ m $ attributes. The predicted label $ \hat y_i $ can be stated as:
\begin{equation}
\hat y_i = \textbf{w}^{T}\textbf{x}_i + \alpha
\end{equation}
where $ (\textbf{x}_i:x_{i1}, x_{i2}, x_{i3}, \cdots, x_{im} )$ are the $ m $ attributes of the $ i $th data record, $ (\textbf{w}: w_{i1}, w_{i2}, w_{i3}, \cdots, w_{im} )$ are the corresponding parameters related to $ m $ attributes, and $\alpha$ is the bias factor of the linear function that helps increase the prediction accuracy.

To measure the error between predicted label and real label, the Euclid distance of a particular data record $ \textbf{x}_i $ is expressed as:
\begin{equation}
d = (\textbf{w}^{T}\textbf{x}_i + \alpha - y_i)^2
\end{equation}
From Eq. (13), we can easily reach the sum of errors over the entire database:
\begin{equation}
(\textbf{w}^*, \alpha^*) = \text{arg min}_{(\textbf{w}, \alpha)} \sum_{i = 1}^{m}(\textbf{w}^{T}\textbf{x}_i + \alpha - y_i)^2
\end{equation}
Eq. (14) is an optimization problem with respect to the parameters $ (\textbf{w}, \alpha) $. In general, the goal of linear regression is to construct a linear function to learn the real label according to attributes of the data sample. 

In real world applications, sometimes people are more interested in the probability prediction for particular tasks. Taking the bankrupt prediction as an example, finance companies would like to learn the bankrupt probability of their customers based on payment history and salary level. Logistic regression is developed to learn the probability for some tasks, and is widely used in various areas to predict the occurrence of particular events, such as incidence of a disease, repurchase probability of a product, and failure rates of facilities. In fact, logistic regression is an extended version of linear regression, but the difference is that we have to map the linear function to a probability prediction. The most commonly used mapping function is sigmoid function written as:  
\begin{equation}
y = \dfrac{1}{1 + e^{-x}}
\end{equation}
Substituting Eq. (12) into Eq. (15), we have:
\begin{equation}
y_i = \dfrac{1}{1 + e^{-(\textbf{w}^{T}\textbf{x}_i + \alpha)}}
\end{equation}
Also, Eq.(16) can be rewritten as:
\begin{equation}
ln\dfrac{y_i}{1-y_i} = {\textbf{w}^{T}\textbf{x}_i + \alpha}
\end{equation}
If we treat $ y_i $ as the probability of positive case and $ 1 - y_i $ as the probability of negative case, and then $ \dfrac{y_i}{1- y_i} $ represents the relative possibility of data sample $ \textbf{x}_i $ to be a positive case. It can be seen in Eq. (16) that logistic regression is to use linear regression results to predict the logarithm probability for occurrence of the real label. 

If we treat $ y_i $ as the poster probability estimation $ p(y_i = 1 | \textbf{x}_i) $, then we can rewrite Eq. (17) as:
\begin{equation}
ln\dfrac{p(y_i = 1 | \textbf{x}_i)}{p(y_i = 0 | \textbf{x}_i)} = \textbf{w}^{T}\textbf{x}_i + \alpha
\end{equation}
It can be derived from Eq. (18) that:
\begin{equation}
{p(y_i = 1 | \textbf{x}_i)} = \dfrac{e^{\textbf{w}^{T}\textbf{x}_i + \alpha}}{1 + e^{\textbf{w}^{T}\textbf{x}_i + \alpha}}
\end{equation}
\begin{equation}
{p(y_i = 0 | \textbf{x}_i)} = \dfrac{1}{1 + e^{\textbf{w}^{T}\textbf{x}_i + \alpha}}
\end{equation}
As a consequence, we can use the maximum likelihood method to estimate $ \textbf{w} $ and $ \alpha $ over the entire dataset \cite{kutner2004applied}:
\begin{equation}
l(\textbf{w}, \alpha) = \sum_{i = 1}^{N}\text{ln} \hspace{0.1cm}p(y_i|\textbf{x}_i; \textbf{w}, \alpha)
\end{equation}
Let $ p_1(\textbf{x};\textbf{w}, \alpha) = p(y = 1 | \textbf{x}; \textbf{w}, \alpha)$, $ p_0(\textbf{x};\textbf{w}, \alpha) = p(y = 0 | \textbf{x}; \textbf{w}, \alpha) = 1 - p_1(\textbf{x};\textbf{w}, \alpha)$. Then the likelihood term in Eq. (21) can be rewritten as:
\begin{equation}
p(y_i | \textbf{x}_i; \textbf{w}, \alpha) = y_ip_1(\textbf{x}_i; \textbf{w}, \alpha) + (1 - y_i)p_0(\textbf{x}_i; \textbf{w}, \alpha)
\end{equation}
Substituting Eq. (22) in to Eq. (21), and according to Eq. (19), Eq. (20), the maximization of Eq. (22) is equivalent to minimizing the following problem:
\begin{equation}
l(\textbf{w}, \alpha) = \sum_{i = 1}^{m}(-y_i(\textbf{w}^T\textbf{x}_i + \alpha) + \text{ln}(1 + e^{\textbf{w}^T\textbf{x}_i + \alpha}))
\end{equation}
The above optimization problem is a differentiable convex function, and can be easily solved by the gradient decent method, or the Newton method.

\subsection{System Architecture and Threat Model}
The system architecture of the collaborative multiparty learning is illustrated in Fig. 1. It is assumed that $ N $ parities are included in this system, and each party has its own local dataset for training. All of the parties agree to train the same model, and logistic regression is applied. The central sever is to maintain the parameters of all these participating parties, including storing, updating, offloading, downloading, and so forth. As we discussed before, each participant constructs a local logistic regression model based on its own dataset. In the initialization step, every participant will obtain a set of parameters $ (\textbf{w}_i, \alpha_i) $, where $ i \in [1, N] $. After obtaining the first round parameters, all these participants will upload their parameters to the central server for sharing. Now the central sever can act as a restoring and exchanging system that allows each participant to download the parameters of others. The advantage is that each participant can use other datasets in learning without knowing the original data of other parties. In addition, each party will not interfere with each other during the training process since the only interaction is the parameter exchanging. It should be noted that weighted parameter sharing is used such that the central server averages the parameters in proportional to the size of each party's data.

\begin{figure}[h]
	\centering{
		\includegraphics[scale=0.44]{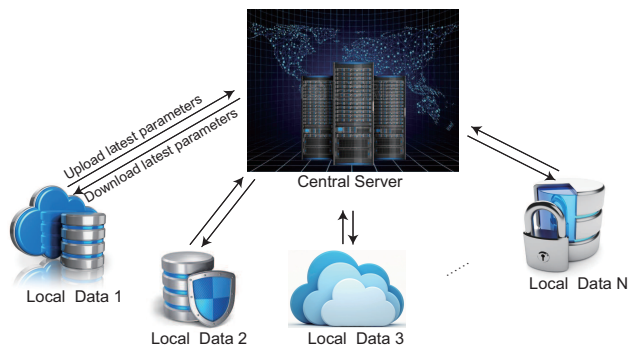}
		\label{fig:Fig1}}
	\caption{System framework for multiparty learning.}\label{fig:issues}
\end{figure}

However, the above multiparty learning system also induces privacy challenges. Since every participant maintains its own dataset which may contain sensitive information,  directly uploading the parameters to the central sever might cause release of private information. As a result, the parameters of each participant should be protected before being sent to the central sever for sharing. In this paper, we will employ two methods to protect the parameters that will ensure differential privacy. The experimental results in section IV will show that both methods can achieve good performance.

\section{Privacy-Preserving Logistic Regression}
\subsection{Output Function Perturbation Approach (OFPA)}\label{sec:ofpa}
As mentioned in the previous section, the parameters $ (\textbf{w}, \alpha) $ may contain sensitive information which cannot be uploaded to the central server directly. Every run of the logistic regression model by the local participant will produce a new set of parameters $ (\textbf{w}, \alpha) $. As we will discuss in Algorithm 1, it is very challenging to directly apply the Laplace mechanism due to the difficulty of calculating the global sensitivity of the objective function. As a result, we choose to develop ways to perturb the objective function and then we can easily apply the Laplace mechanism to preserve differential privacy. Let us first start with a simple approach which directly adds Laplace noise to the parameters $ (\textbf{w}, \alpha) $ [13], as described in \textit{Algorithm 1}.  
\begin{algorithm}[]   
	\caption{ Laplace noise addition to parameters}   
	\label{alg:Framwork1}   
	\begin{algorithmic}[1] 
		\REQUIRE ~~\\ 
		Input a local database  $ \mathcal{D}$; \\
		\ENSURE ~~\\ %算法的输出：Output 
		Encrypted parameter $ (\textbf{w}, \alpha) $;   
		\STATE Construct a local logistic regression model $ \mathcal{M} $ over the database $ \mathcal{D} $;
		\STATE Compute the optimal parameter $ (\textbf{w}^*, \alpha^*) $ for $ \mathcal{M} $;
		\STATE Compute $ \Delta S_{\mathcal{M}} $ from Eq. (6);
		\STATE Generate a random noise vector \textbf{v} with elements from the Laplace distribution $ Lap(x|0, \dfrac{\Delta S_{\mathcal{M}}}{\epsilon}) = \dfrac{\epsilon}{2\Delta S_{\mathcal{M}}}exp(-\dfrac{|x|\epsilon}{\Delta S_{\mathcal{M}}}) $;
		\STATE Compute $ (\textbf{w}, \alpha) = (\textbf{w}^*, \alpha^*) + \textbf{v}$; 
		\RETURN $ (\textbf{w}, \alpha) $;  
	\end{algorithmic}  
\end{algorithm}

\textbf{Theorem 1.} The output $ (\textbf{w}, \alpha) = (\textbf{w}^*, \alpha^*) + \textbf{v}$ from \textit{Algorithm 1} satisfies $\epsilon$-differential privacy. 
\\
\\ \textit{Proof}: The proof is omitted due to space limit. 
\\
\\
\indent From \textbf{Theorem 1}, we know that \textit{Algorithm 1} provides a way to realize $ \epsilon $-differential privacy with respect to parameters $ (\textbf{w}, \alpha) $ of each local participant. However, the disadvantage of \textit{Algorithm 1} is the complexity of computing global sensitivity $ \Delta S_{\mathcal{M}} $, which may not be preferable for model training. 
\\ \indent Next we will develop a more brief and stable method to realize $ \epsilon $-differential privacy for the parameters $ (\textbf{w}, \alpha) $. Instead of adding noise to the parameters directly, we decide to generate a noise vector $ \textbf{v} $ via the Laplace mechanism, and adding this generated noisy vector into objective function which will give:
\begin{equation}
\begin{split}
l'(\textbf{w},\alpha) &= l(\textbf{w}, \alpha) + \textbf{v}^T\textbf{w} 
\\ & = \sum_{i = 1}^{m}(-y_i(\textbf{w}^T\textbf{x}_i + \alpha) + ln(1 + e^{\textbf{w}^T\textbf{x}_i + \alpha})) + \textbf{v}^T\textbf{w}
\end{split}
\end{equation}
It can be proved that the output of Eq. (24) meets $ \epsilon $-differential privacy as stated in \textbf{Theorem 2}. The detailed computational procedures for collaborative learning of OFPA is shown in \textit{Algorithm 2}.   

The $ 3 $rd line of \textit{Algorithm 2} points out the stop criteria for collaborative learning. If the variation at the central server is below the preset threshold, we assume the result is optimal and the parameters obtained are global optimal for all participants. In the $ 7 $th line, it can be seen that for every participant's objective function, we will add a Laplace noise vector to protect the privacy. The $ 8 $th line is the core part of collaborative learning that every participant will download the weighted averaging parameters for the next-round training. The main function of the central server is stated in line $ 12 $; i.e., it will calculate the weighted average of the uploaded parameters from all participants. From the whole process, participants can enjoy the benefits of multiple data sources without worrying about information leakage.
\\
\textbf{Theorem 2.} The output of $ l'(\textbf{w}, \alpha) = l(\textbf{w}, \alpha) + \textbf{v}_{ik}^T\textbf{w}$ for each participant in \textit{Algorithm 2} satisfies $\epsilon$-differential privacy.
\\
\\\textit{Proof}: To prove the $\epsilon$-differential privacy for any pair of neighbor databases $ \mathcal{D} $ and $ \mathcal{D'} $, we need to show that \textit{Definition 1} holds for any randomized algorithms. %From \textit{Definition 1}, we know that $ \mathcal{D} $ and $ \mathcal{D'} $ are differing at only one item. 
\\ \indent Without loss of generality, suppose the last element of $ \mathcal{D} $ and $ \mathcal{D'} $ is different, such that $ \mathcal{D} $ is composed of $ (\textbf{x}_1, y_1), (\textbf{x}_2, y_2), \cdots , (\textbf{x}_{n-1}, y_{n-1}), (\textbf{x}_n, y_n) $, and  $ \mathcal{D'} $ is composed of $ (\textbf{x}_1, y_1), (\textbf{x}_2, y_2), \cdots $, $(\textbf{x}_{n-1}, y_{n-1}), (\textbf{x}'_n, y'_n) $. In addition, we also assume that $ \parallel\textbf{x}_i\parallel \leq 1$, which can be normalized to $ 1 $ if not. We know that the minimization of Eq. (24) will lead to the zero derivative. In addition, let $ (\textbf{w}, \alpha) $ and $ (\textbf{w}', \alpha') $ be what are obtained for $ \mathcal{D} $ and $ \mathcal{D'} $ after every round of training.
\\ \indent To prove $\epsilon$-differential privacy between $ \mathcal{D} $ and $ \mathcal{D'} $, we only need to show $\epsilon$-differential privacy of the output from $ \l'(\textbf{w}, \alpha)$. The zero derivative of Eq. (24) for the last element of $ \mathcal{D} $ and $ \mathcal{D'} $ will give:
\begin{equation}
\begin{split}
&\textbf{v} - y_n\textbf{x}_n - \dfrac{\textbf{x}_n}{1 + exp(\textbf{w}_n^T\textbf{x}_n + \alpha)} \\&= \textbf{v}' - y_n\textbf{x}_n' - \dfrac{\textbf{x}_n'}{1 + exp(\textbf{w}_n^{T'}\textbf{x}'_n + \alpha')}
\end{split}
\end{equation}
Since $ \parallel \textbf{x}_n \parallel_1 \leq 1$, $ \parallel \textbf{x}'_n \parallel_1 \leq 1$, $exp(\textbf{w}_n^{T}\textbf{x}_n + \alpha) > 0$, and $ exp(\textbf{w}_n^{T'}\textbf{x}'_n + \alpha') > 0$. Thus for any pair of $ \textbf{v} $ and $ \textbf{v}' $, we will get $ \parallel \textbf{v} - \textbf{v}' \parallel_1 \leq 4$ from Eq. (25). Then for any $ (\textbf{w}, \alpha) $ we have the following:
\begin{equation}
\begin{split}
&\dfrac{P((\textbf{w},\alpha)|(\textbf{x}_1, y_1), (\textbf{x}_2, y_2), (\textbf{x}_3, y_3), \cdots , (\textbf{x}_n, y_n))}{P((\textbf{w}, \alpha)|(\textbf{x}_1, y_1), (\textbf{x}_2, y_2), (\textbf{x}_3, y_3), \cdots , (\textbf{x}'_n, y'_n))}\\ &= \dfrac{\Pi_{i=1}^{m}e^{\frac{-4}{\epsilon} v_i}}{\Pi_{i=1}^{m}e^{\frac{-4}{\epsilon} v_i'}}
\\ &= e^{\dfrac{-4}{\epsilon}\parallel \textbf{v} \parallel_1- \parallel \textbf{v}' \parallel_1} \leq e^{\epsilon}
\end{split}
\end{equation}
\begin{algorithm}[h]   
	\caption{ Privacy-Preserving Collaborative Logistic Regression of OFPA}   
	\label{alg:Framwork2}   
	\begin{algorithmic}[1] 
		\REQUIRE ~~\\ 
		Input $ N $ databases $ \mathcal{D}_1, \mathcal{D}_2, \mathcal{D}_3, \cdots, \mathcal{D}_N$; \\
		\ENSURE ~~\\ Output global optimal parameter $ (\textbf{w}, \alpha) $; 
		\STATE Initialize \textit{k  = 0};
		\STATE Initialize $(\textbf{w}_k, \alpha_k) = \textbf{0}$;
		\STATE \textbf{while} $ \mid \Delta(\textbf{w}_k, \alpha_k) \mid > \eta$   
		\STATE \hspace{0.5cm}\textbf{for} i = 1 : $ N $
		\STATE \hspace{1.0cm}Local data owner $ i $ constructs its own logistic re-\\ \hspace{1.0cm}gression model $ l(\textbf{w},\alpha) $ from Eq. (23);
		\STATE \hspace{0.9cm} Generate a random vector $ \textbf{v}_{ik} $ with elements from \\ \hspace{1.0cm}Laplace distribution $ Lap(x|0, 4/{\epsilon})$;
		\STATE \hspace{1.0cm}Compute perturbed objective function $ l'(\textbf{w}, \alpha) = $ \\$ \hspace{1.0cm}l(\textbf{w}, \alpha) + \textbf{v}_{ik}^T\textbf{w}$; 
		\STATE \hspace{0.9cm} Download weighted parameters from central 
		\\ \hspace{1.0cm}server, and update $ (\textbf{w}_{ik}, \alpha_{ik}) = (\textbf{w}_{k}, \alpha_{k}) $;
		\STATE \hspace{0.9cm} Compute $ (\textbf{w}^{*}_{ik}, \alpha^{*}_{ik}) = \text{arg min}\hspace{0.1cm} l'(\textbf{w}, \alpha)$;
		\STATE \hspace{0.9cm} Upload $ (\textbf{w}^*_{ik}, \alpha^*_{ik}) $ to the central server;
		\STATE \hspace{0.4cm} \textbf{end for};
		\STATE \hspace{0.4cm} Central server computes weighted averaging param- \\ \hspace{0.42cm} eter $ (\textbf{w}_{k}, \alpha_{k}) $;
		\STATE \hspace{0.4cm} $ k = k + 1 $;
		\RETURN $ (\textbf{w}, \alpha) = (\textbf{w}_{k}, \alpha_{k})$; 
	\end{algorithmic}  
\end{algorithm}

\subsection{Output Function Approximation Approach (OFAA)}\label{sec:ofaa}
For OFPA, we add a noise vector into the objective function according to Laplace mechanism. However, one can see that the generated noise is from a constant scale Laplace distribution, and we cannot adjust the noise level. As a result, a preferable mechanism should be able to adjust the noise level according to particular forms of the objective function. In the following, we will develop an approach by  injecting noise to coefficients of the objective function's approximation form. 

Before delving into details of the function approximation approach, we first discuss the structure of the objective function of logistic regression. It can be verified that $ l(\textbf{w}, \alpha) $ in Eq. (23) is a continuous and differentiable function. According to Stone–Weierstrass Theorem \cite{rudin1964principles}, we can approximate $ l(\textbf{w}, \alpha) $ with a polynomial function with respect to $ (\textbf{w}, \alpha) $. Parameter $ (\textbf{w}, \alpha) $ is a $ m + 1 $ dimensional vector variable $ (w_1, w_2, w_3, \cdots, w_m, \alpha) $. Let $ \Phi_j $ be the set of products of $ (w_1, w_2, w_3, \cdots, w_m, \alpha) $ at the $ j $th degree expressed as:
\begin{equation}
\Phi_j = \{ w_1^{d_1}\cdot w_2^{d_2} \cdot w_3^{d_3} \cdots w_m^{d_m}\cdot \alpha^{d_{ m + 1}} | \sum_{i = 1}^{m+ 1} d_i = j \}
\end{equation}
where $ d_1, d_2, d_3, \cdots, d_{m + 1} \in \mathcal{N} $, and let $ \phi(\textbf{w}) =  w_1^{d_1}\cdot w_2^{d_2} \cdot w_3^{d_3} \cdots w_m^{d_m}\cdot \alpha^{d_{ m + 1}}$. Then we can get the approximation expression of  $ l(\textbf{w}, \alpha) $ according to Stone–Weierstrass Theorem as follows:
\begin{equation}
l(\textbf{w}, \alpha) = \sum_{j = 0}^{J} \sum_{\phi \in \Phi_j} \sum_{s_i \in \mathcal{D}} \lambda_{\phi s_i} \phi(\textbf{w})
\end{equation}
where $s_i$ is the $i$th element in database $ \mathcal{D} $, and $ \lambda_{\phi s_i} $ is the coefficient of polynomial for data record $ s_i $. It can be seen from Eq. (28) that the objective function $ l(\textbf{w}, \alpha)  $ can be approximated with a formula consisting of polynomial function only. As a result, it occurs to us that we can add noise to the coefficients of each degree in the polynomial form. The following \textit{Algorithm} 3 will give detailed steps about adding noise to the coefficients of the polynomial expression. 

From \textit{Algorithm 3}, we can see that the central server plays the same role of the previous algorithm, but the difference lies in the noise generation mechanism. The determination of the noise level is from the $7$th and $8$th steps of \textit{Algorithm 3}. For the $j$th degree of the approximate polynomial expression, we choose the maximum coefficient $ \parallel \lambda_{\Phi_j}\parallel_{max} $, and set $ \Delta S = 2(J + 1)\parallel \lambda_{\Phi_j}\parallel_{max} $ as the scale for Laplace distribution. Then in the $10$th step, we will obtain a new perturbed approximate objective function $ \hat l(\textbf{w}, \alpha) $. The parameter $(\textbf{w}, \alpha)$ from $ \hat l(\textbf{w}, \alpha) $ satisfies $ \epsilon$-differential privacy, and the proofs will be given in \textbf{Theorem 3}. 
\\
\\ \textbf{Theorem 3}. In \textit{Algorithm 3}, the perturbed approximate objective function $ \hat l(\textbf{w}, \alpha) $ satisfies $ \epsilon$-differential privacy for Laplace distribution with scale $ \Delta S $. 
\\ 
\\\textit{Proof:} Firstly, without loss of generality, we suppose $ \mathcal{D} $ and $\mathcal{D'}$ differ with the last data record, and let $ s_n $ and $s_n'$ be the corresponding last data records. We have that:
\begin{equation}
\begin{split}
&\dfrac{P((\textbf{w},\alpha)|(s_1, s_2, s_3, \cdots , s_n) }{P((\textbf{w},\alpha)|(s_1, s_2, s_3, \cdots , s_n') }
\\ &= \dfrac{\Pi_{j = 0}^{J} \hspace{0.1cm}\Pi_{\phi \in \Phi_j}\hspace{0.1cm}\Pi_{s_i \in \mathcal{D}}\hspace{0.1cm}e^{(\dfrac{\epsilon\parallel \lambda_{\phi s_i}' - \lambda_{\phi s_i} \parallel_1}{\Delta S})}}{\Pi_{j = 0}^{J} \hspace{0.1cm}\Pi_{\phi \in \Phi_j}\hspace{0.1cm}\Pi_{s_i \in \mathcal{D'}}\hspace{0.1cm}e^{(\dfrac{\epsilon\parallel \lambda_{\phi s_i}' - \lambda_{\phi s_i} \parallel_1}{\Delta S})}}
\\ &\leq \Pi_{j = 0}^{J} \hspace{0.1cm}\Pi_{\phi \in \Phi_j}\hspace{0.1cm}e^{(\dfrac{\epsilon(\parallel\sum_{s_i \in \mathcal{D}} \lambda_{\phi s_i}' - \sum_{s_i \in \mathcal{D'}} \lambda_{\phi s_i}' \parallel_1)}{\Delta S}})
\\ & =\Pi_{j = 0}^{J} \hspace{0.1cm}\Pi_{\phi \in \Phi_j}\hspace{0.1cm}e^{(\dfrac{\epsilon(\parallel \lambda_{\phi s_n}'  -  \lambda_{\phi s_n'}' \parallel_1)}{\Delta S}})
\\ & = \Pi_{j = 0}^{J} \hspace{0.1cm} e^{(\dfrac{\epsilon(\parallel \lambda_{\Phi_j s_n}'  -  \lambda_{\Phi_j s_n'}' \parallel_1)}{\Delta S}}
\\ & \leq \Pi_{j = 0}^{J} \hspace{0.1cm} e^{(\dfrac{\epsilon(2 \hspace{0.1cm}\text{max}(\parallel \lambda_{\Phi_j}' \parallel_{1}, \parallel \lambda_{\Phi_j }' \parallel_{1}))}{\Delta S}}
\\ & = e^{\sum_{j = 0}^{J}\epsilon/ (J + 1)} = e^{\epsilon}
\end{split}
\end{equation}
The first inequity is derived from the triangle formula $ \parallel a \parallel_1 - \parallel b \parallel_1 \hspace{0.1cm}\leq \hspace{0.1cm}\parallel a \pm b \parallel_1$, where $ a $ and $b$ are real numbers. The second inequality is derived as follows:
\begin{equation}
\begin{split}
\parallel \lambda_{\Phi_j s_n}'  -  \lambda_{\Phi_j s_n'}' \parallel_1 &\leq \parallel \lambda_{\Phi_j s_n}'\parallel  + \parallel \lambda_{\Phi_j s_n'}' \parallel_1 \nonumber
\\& \leq 2 \hspace{0.1cm}\text{max}(\parallel \lambda_{\Phi_j}' \parallel_{1}, \parallel \lambda_{\Phi_j}' \parallel_{1}) 
\end{split}
\end{equation}
Then Eq.(29) holds, and the proof is complete.
\begin{algorithm}[h]   
	\caption{ Privacy-Preserving Collaborative Logistic Regression of OFAA}   
	\label{alg:Framwork}   
	\begin{algorithmic}[1] 
		\REQUIRE ~~\\ 
		Input $ N $ databases $ \mathcal{D}_1, \mathcal{D}_2, \mathcal{D}_3, \cdots, \mathcal{D}_N$; \\
		\ENSURE ~~\\ Output global optimal parameter $ (\textbf{w}, \alpha) $; 
		\STATE Initialize \textit{k  = 0};
		\STATE Initialize $(\textbf{w}_k, \alpha_k) = \textbf{0}$;
		\STATE \textbf{while} $ \mid \Delta(\textbf{w}_k, \alpha_k) \mid > \eta$   
		\STATE \hspace{0.3cm} \textbf{for} i = 1 : $ N $
		\STATE \hspace{0.6cm} Let $ l(\textbf{w}, \alpha)_i = \sum_{j = 0}^{J} \sum_{\phi \in \Phi_j} \sum_{s_k \in \mathcal{D}_i} \lambda_{\phi s_k} \phi(\textbf{w},\alpha) $;
		\STATE \hspace{0.6cm} \textbf{for} j = 0 : $ J $
		\STATE \hspace{1.1cm} Let $\Delta S = 2(J + 1)\hspace{0.05cm} \hspace{0.05cm}\parallel \lambda_{\Phi_j}\parallel_{max}$;
		\STATE \hspace{1.1cm} Let $ \lambda_{\Phi_j}' = \lambda_{\Phi_j} + Lap(x|0, \dfrac{\Delta S}{\epsilon})$;
		\STATE \hspace{0.6cm} \textbf{end for};
		\STATE \hspace{0.6cm} Let $ \hat l(\textbf{w}, \alpha)_i = \sum_{j = 0}^{J} \sum_{\phi \in \Phi_j} \sum_{s_k \in \mathcal{D}_i} \lambda_{\phi' s_k} \phi(\textbf{w}, \alpha) $;
		\STATE \hspace{0.6cm} Download weighted parameters from central server, \\ \hspace{0.75cm}and update $ (\textbf{w}_{ik}, \alpha_{ik}) = (\textbf{w}_{k}, \alpha_{k}) $;
		\STATE \hspace{0.6cm} Compute $ (\textbf{w}^*_{ik}, \alpha^*_{ik}) = \text{arg min} \hspace{0.1cm} \hat l(\textbf{w}, \alpha)_i$;
		\STATE \hspace{0.6cm} Upload $ (\textbf{w}^*_{ik}, \alpha^*_{ik})$ to central server;
		\STATE \hspace{0.3cm} \textbf{end for};
		\STATE \hspace{0.3cm} Central server computes weighted averaging param- \\ \hspace{0.32cm} eter $ (\textbf{w}_{k}, \alpha_{k}) $;
		\STATE \hspace{0.3cm} $ k = k + 1 $;
		\RETURN $ (\textbf{w}, \alpha) = (\textbf{w}_k, \alpha_k)$; 
	\end{algorithmic}  
\end{algorithm}
\\ 

It can be seen that the noise addition mechanism of \textit{Algorithm 3} guarantees $\epsilon$-differential privacy. As a result, releasing the parameter of $ \hat l(\textbf{w}, \alpha)$ will not cause information leakage. In addition, the noise addition mechanism in \textit{Algorithm 3} is designed for objective function with polynomial expression, but $ l(\textbf{w}, \alpha) $ is not of polynomial form yet. We have stated that $ l(\textbf{w}, \alpha) $ can be approximated with a polynomial form by the Stone–Weierstrass Theorem and next we will find a way to derive the approximation polynomial form of  $ l(\textbf{w}, \alpha) $. 

It is well known that Taylor expansion is commonly used in approximating a continuous and differentiable function with arbitrary precision. In this paper, we decide to use Taylor expansion to derive the approximation polynomial expression for our objective function $ l(\textbf{w}, \alpha) $. For convenience, we can rewrite $ l(\textbf{w}, \alpha) $ as the following:
\begin{equation}
l(\textbf{w}, \alpha) = \sum_{i = 1}^{m}[l_1(\theta_1)- l_2(\theta_2)]
\end{equation}
where $ l_1(t) = \text{ln}(1 + \text{exp}(t)) $, $ l_2(t) = t $, $ \theta_1 = \textbf{w}^T\textbf{x}_i + \alpha$, and $ \theta_2 =y_i( \textbf{w}^T\textbf{x}_i + \alpha)$. In addition, the Taylor expansion of a differentiable and continuous function $ f(x) $ at point $ a $ is expressed as:
\begin{equation}
\sum_{n = 0}^{\infty}\dfrac{f^{(n)}(a)}{n!}(x - a)^n
\end{equation}
where $ f^{(n)}(a) $ is the $ n $th derivative of function $ f $ evaluated at point $ a $. If we apply Taylor expansion to $ l(\textbf{w}, \alpha) $, we will have the following:
\begin{equation}
l(\textbf{w}, \alpha) = \sum_{i = 1}^{m}\sum_{n = 0}^{\infty} [\dfrac{l_1^{(n)}(\theta_1')}{n!}(\theta_1 - \theta_1')^n - \dfrac{l_2^{(n)}(\theta_2')}{n!}(\theta_2 - \theta_2')^n]
\end{equation}
where $ l_1^{(n)}(\theta_1') $ is the $ n $th derivative of function $ \theta_1 $ evaluated at point $ \theta_1' $, and $ l_2^{(n)}(\theta_2') $ is the $ n $th derivative of function $ \theta_2 $ evaluated at point $ \theta_2' $. Note that $ l_2^{(n)} = 0 $ for $ n > 1 $. As a result, we can simplify Eq.(32) by setting $ \theta_1' = \theta_2' = 0 $, and we will have the following:
\begin{equation}
l(\textbf{w}, \alpha) = \sum_{i = 1}^{m}\sum_{n = 0}^{\infty} \dfrac{l_1^{(n)}(0)}{n!}(\textbf{w}^T\textbf{x}_i + \alpha)^n - \sum_{i = 1}^{m}(y_i\textbf{w}^T\textbf{x}_i + \alpha)
\end{equation}
Up to now, we have derived the polynomial expression of the objective function $l(\textbf{w}, \alpha)$. However, we cannot apply Eq. (33) directly due to the infinite summation. We can remove higher order polynomial terms of the Taylor expansion and only keep terms with orders $ n \leq 2 $. Then the approximate polynomial form of Eq. (33) is expressed as:
\begin{equation}
l(\textbf{w}, \alpha) = \sum_{i = 1}^{m}\sum_{n = 0}^{2} \dfrac{l_1^{(n)}(0)}{n!}(\textbf{w}^T\textbf{x}_i + \alpha)^n - \sum_{i = 1}^{m}(y_i\textbf{w}^T\textbf{x}_i + \alpha)
\end{equation}
and calculation of the derivative shows that $ l_1^{(0)}(0) = \text{ln}2 $, $ l_1^{(1)}(0) = 1/2 $, and $ l_2^{(0)}(0) = 1/4 $. With these derivative results, we can calculate the scale in \textit{Algorithm 3}. As described in \textit{Algorithm 3}, for the polynomial expression with different order $ j $, we will choose the according $ \Delta S = 2(J + 1)\parallel \lambda_{\Phi_j}\parallel_{max} $ as the scale of the Laplace distribution. Taking $ j = 1 $ as an example, $ \Delta S $ is expressed as:
\begin{equation}
\begin{split}
\Delta S &=  2 \hspace{0.1cm}\text{max} (\dfrac{f_1^{0}(0)}{1!}\sum_{i = 1}^{d}x_i + \alpha + y_i\sum_{i = 1}^{d}x_i + \alpha)*3
\\ & \leq \dfrac{9}{2}d
\end{split}
\end{equation}
where $ d $ is the number of attributes of data records. As a result, we can inject the noise into the coefficients of the first order polynomial term with Laplace distribution $ Lap(x|0, \dfrac{9d}{2\epsilon}) $ according to \textit{Algorithm 3}, and inject noise to the polynomial form of other orders with the same approach.

\section{Experiments}
In this section, we evaluate the performance of the two proposed approaches OFPA and OFAA, and compare them with regular logistic regression without privacy protection denoted as \textit{LR\_NoPrivacy} which is trained as a single party who holds the entire dataset. All experiments are conducted using Python 2.7 on a Macbook with a 2.2 GHz Intel Core i7 CPU and 16GB RAM.

We choose two real datasets for experiments, \textit{Bank Marketing} \cite{moro2014data} used to predict whether the client will subscribe a term deposit and \textit{Default of Credit Card Clients} \cite{yeh2009comparisons} for predicting whether a credit owner will default or not. The \textit{Bank Marketing} dataset contains 45211 records and 17 attributes, including bank client attributes (e.g., age), current campaign attributes (e.g., contact communication type), and social and economic context attributes (e.g., employment variation rate). For the \textit{Default of Credit Card Clients} dataset, it contains 30000 records and 24 attributes, including credit card owner attributes (e.g., gender), history of past payment, amount of bill statement attributes, and amount of previous payment attributes. Since several attributes in both datasets are categorical variables, we encode such categorical values into integers for our model training using \textit{LabelEncoder} \cite{labelencoder}. Without loss of generality, we assume three parties are included intending to collaboratively learn the logistic regression model, which hold 40\%, 30\% and 10\% of each dataset separately, while the remained 20\% of each dataset is treated as test set.

We conduct logistic regression on each dataset by varying three different parameters, namely the privacy budget $\epsilon$, dataset cardinality, and dataset dimensionality.  According to Eq. (19), if ${p(y_i = 1 | \textbf{x}_i)} = \dfrac{e^{\textbf{w}^{T}\textbf{x}_i + \alpha}}{1 + e^{\textbf{w}^{T}\textbf{x}_i + \alpha}} > 0.5$, we make prediction to be 1, and otherwise to be 0. The accuracy of logistic regression models is measured by \textit{misclassification rate}, which is defined as the fraction of data records that are incorrectly classified. Additionally, in each experiment, the logistic regression model is trained 40 epochs 10 times, and the average results are reported.  

\subsection{Classification Accuracy vs. Privacy Budget}
To explore how privacy budget $\epsilon$ affects the performance of the proposed algorithms, experiments are performed by varying $\epsilon$ from 0.1 to 3.2. Fig. 2 shows the misclassification rate of each algorithm against the privacy budget $\epsilon$. The accuracy of LR\_NoPrivacy almost stays stable on both datasets. Both OFPA and OFAA produce less misclassifications with increasing  $\epsilon$, since a larger $\epsilon$ means that a smaller amount of noise is added to the objective function. Furthermore, it shows that the performance of OFPA is slightly better than OFAA, but both of them are robust against the varied $\epsilon$ and close to regular logistic regression.

\begin{figure}[h]
	\centering{
		\includegraphics[scale=0.4]{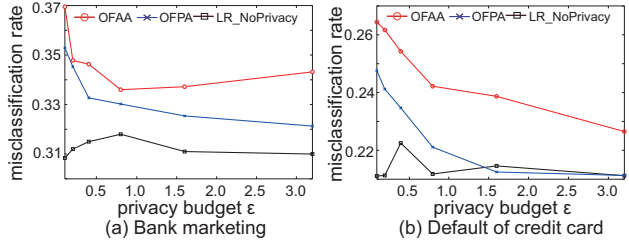}
		\label{fig:budget}}
	\caption{Classification Accuracy vs. Privacy Budget.}\label{fig:issues}
\end{figure}

\subsection{Classification Accuracy vs. Dataset Cardinality}
To evaluate the classification accuracy against the variation of dataset cardinality, we generate random subsets of the two original datasets with sampling rate from 0.2 to 1, while keeping $\epsilon$ as 0.8. As shown in Fig. 3, the accuracy of LR\_NoPrivacy slightly outperforms that of OFPA and OFAA, but the performance of OFPA and OFAA improves rapidly with the increase of dataset cardinality. More importantly, the misclassification rate of OFPA and OFAA is comparable with LR\_NoPrivacy when we use the full dataset, indicating that our proposed algorithms can make accurate predictions while protecting data privacy.

\begin{figure}[h]
	\centering{
		\includegraphics[scale=0.4]{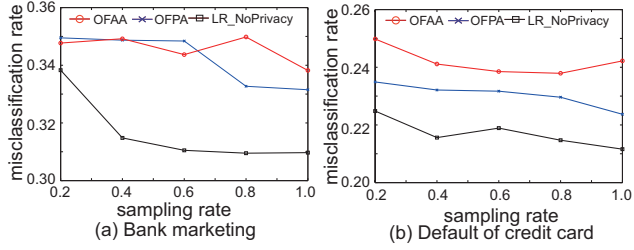}
		\label{fig:sample}}
	\caption{Classification Accuracy vs. Dataset Cardinality.}\label{fig:issues}
\end{figure}

\subsection{Classification Accuracy vs. Dataset Dimensionality}
To demonstrate the effectiveness of the proposed algorithms against the change of dataset dimensionality, we vary the dimensions of \textit{Bank Marketing} dataset from 5 to 17, and change the dimensionality of \textit{Default of Credit Card Clients} from 8 to 24, while setting $\epsilon$ as 0.8. Fig. 4 shows that although the performance of LR\_NoPrivacy still performs slightly better than that of OFPA and OFAA, the accuracy of our proposed approaches improves with the increasing of dataset dimensionality. %In addition, when we include all dimensions of original datasets, the accuracy of both OFPA and OFAA are almost identical with that of LR\_NoPrivacy on \textit{Bank Marketing}, and the accuracy of OFPA on \textit{Default of Credit Card Clients} is similar to that of LR\_NoPrivacy.

\begin{figure}[h]
	\centering{
		\includegraphics[scale=0.4]{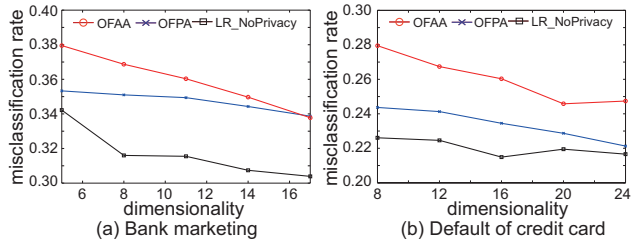}
		\label{fig:dimensionality}}
	\caption{Classification Accuracy vs. Dataset Dimensionality.}\label{fig:issues}
\end{figure}

\subsection{Training Time vs. Privacy Budget}
In order to evaluate how noise injection affects the training time, we train each logistic regression model with 40 epochs 10 times. The average training time is reported in Fig. 5. It shows that time cost of training LR\_NoPrivacy is less than that of OFPA and OFAA, which is reasonable since the latter needs more time to stabilize the noisy model. Note that the training of logistic model using OFAA is slower than that of OFPA, since the injected noise in OFPA is constant, but we have to generate more noise to perform OFAA.

\begin{figure}[h]
	\centering{
		\includegraphics[scale=0.4]{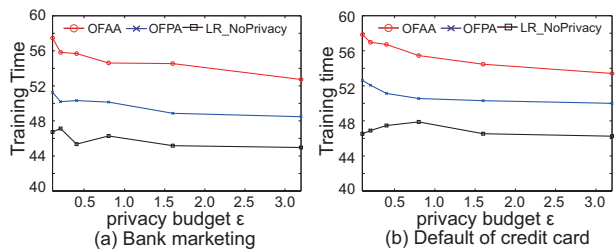}
		\label{fig:train}}
	\caption{Training Time vs. Privacy Budget.}\label{fig:issues}
\end{figure}

\section{Related Work}
Dwork et al. first proposed the notion of $ \epsilon $-differential privacy \cite{dwork2008differential}, and provided Laplace mechanism to achieve it. Later, differential privacy has been developed as a platform to deal with privacy analysis and extensive work employed it to address different types of tasks. For example, Friedman et al. \cite{friedman2010data} achieved $\epsilon $-differential privacy decision tree to predict adult incomes. Raghav et al. \cite{bhaskar2010discovering} reported a differential privacy solution for frequent pattern mining. Other types of work preserving differential privacy have also been done, including support vector machines \cite{li2014privacy}, recommendation system \cite{mcsherry2009differentially}, and neural networks \cite{abadi2016deep}. Differential privacy related to logistic regression has also been demonstrated. For instance, Chaudhuri et al. \cite{chaudhuri2009privacy} enforced $ \epsilon $-differential privacy for logistic regression analysis, but the cost function considered is not of standard regression form. Zhang et al. \cite{zhang2012functional} proposed to approximate the objective function while adding noise to it; however, the addition noise level is high because noise is determined by the global approximation form. In general, all of the above work focus on single party training without considering the multiparty setting.

Some research work has been done on privacy-preserving learning from multiparty data. Pathak et al. \cite{pathak2010multiparty} proposed a differentially private algorithm based on parameter averaging through secure multiparty computation. Rajkumar et al. \cite{rajkumar2012differentially} designed a privacy-preserving multiparty learning scheme, which is enforced by private exchange of gradient information to minimize empirical risks incrementally. In addition, other works using different forms of noise scaling to achieve differential privacy over distributed data have also been reported \cite{shokri2015privacy,heikkila2017differentially}. Different from previous work, we propose a weighted sharing scheme which will help increase the accuracy of model. In addition, we propose to approximate the objection and then inject noise to each degree separately, providing a more efficient, more concise and faster method to complete data training. Furthermore, the schemes designed here feature easy extension to other machine learning tasks. 

\section{Conclusion}
In this paper, we proposed two differentially private approaches for collaboratively training logistic regression classifiers among multiple parties. The proposed approaches enable users to enjoy well-trained logistic regression classifiers based on distributed datasets without disclosing their raw data to each other. Experimental results show that the effectiveness, robustness, and training cost of the proposed algorithms are close to that of regular logistic regression on the aggregate dataset without privacy protection. Although this work mainly focuses on logistic regression, the proposed schemes can also be extended to other classification problems in the context of collaborative learning.\vspace{-0.1in}

\bibliographystyle{splncs03}
\bibliography{INFOCOM2018}

\end{document}